\documentclass[twocolumn]{aastex62}

\usepackage{bm}
\usepackage{natbib,bm,multirow,tabularx}
\shortauthors{Chung et al.}
\shorttitle{Discovery of a wide orbit Jupiter-mass planet}

\newcommand{\te}{t_{\rm E}}
\newcommand{\tzero}{t_{0}}
\newcommand{\uo}{u_{0}}

\newcommand{\fss}{f_{\rm s}}
\newcommand{\fb}{f_{\rm b}}
\newcommand{\pie}{\pi_{\rm E}}
\newcommand{\pien}{\pi_{\rm E,N}}
\newcommand{\piee}{\pi_{\rm E,E}}
\newcommand{\bpie}{\bm{\pi}_{\rm E}}

\newcommand{\thetae}{\theta_{\rm E}}
\newcommand{\thetas}{\theta_{\rm \star}}

\newcommand{\murel}{\mu_{\rm rel}}
\newcommand{\dl}{D_{\rm L}}
\newcommand{\ds}{D_{\rm S}}
\newcommand{\delcs}{\Delta \chi^{2}}
\newcommand{\thisevent}{OGLE-2019-BLG-1180}

\begin{document}

\title{\thisevent Lb: Discovery of a Wide-orbit Jupiter-mass planet around a late-type star}

\correspondingauthor{Sun-Ju Chung}
\email{sjchung@kasi.re.kr}

\author{Sun-Ju Chung}
\affiliation{Korea Astronomy and Space Science Institute, 776 Daedeokdae-ro, Yuseong-Gu, Daejeon 34055, Republic of Korea}
\affiliation{Center for Astrophysics $|$ Harvard \& Smithsonian, 60 Garden St., Cambridge, MA 02138, USA} 

\author{Andrzej Udalski}
\affiliation{Astronomical Observatory, University of Warsaw, AI.~Ujazdowskie~4, 00-478~Warszawa, Poland}

\author{Jennifer C. Yee}
\affiliation{Center for Astrophysics $|$ Harvard \& Smithsonian, 60 Garden St., Cambridge, MA 02138, USA} 

\author{Andrew Gould}
\affiliation{Department of Astronomy, Ohio State University, 140 W. 18th Avenue, Columbus, OH 43210, USA}
\affiliation{Max-Planck-Institute for Astronomy, Königstuhl 17, D-69117 Heidelberg, Germany}
\collaboration{(Leading Authors),}

\author{Michael D. Albrow}
\affiliation{Department of Physics and Astronomy, University of Canterbury, Private Bag, 4800 Christchurch, New Zealand}

\author{Youn Kil Jung}
\affiliation{Korea Astronomy and Space Science Institute, 776 Daedeokdae-ro, Yuseong-Gu, Daejeon 34055, Republic of Korea}
\affiliation{University of Science and Technology, Korea, (UST), 217 Gajeong-ro, Yuseong-gu, Daejeon 34113, Republic of Korea}

\author{Kyu-Ha Hwang}
\affiliation{Korea Astronomy and Space Science Institute, 776 Daedeokdae-ro, Yuseong-Gu, Daejeon 34055, Republic of Korea}

\author{Cheongho Han}
\affiliation{Department of Physics, Chungbuk National University, Cheongju 361-763, Republic of  Korea}

\author{Yoon-Hyun Ryu}
\affiliation{Korea Astronomy and Space Science Institute, 776 Daedeokdae-ro, Yuseong-Gu, Daejeon 34055, Republic of Korea}

\author{In-Gu Shin}
\affiliation{Center for Astrophysics $|$ Harvard \& Smithsonian, 60 Garden St., Cambridge, MA 02138, USA}

\author{Yossi Shvartzvald}
\affiliation{Department of Particle Physics and Astrophysics, Weizmann Institute of Science, Rehovot 76100, Israel}

\author{Hongjing Yang}
\affiliation{Department of Astronomy and Tsinghua Centre for Astrophysics, Tsinghua University, Beijing 100084, China}

\author{Weicheng Zang}
\affiliation{Center for Astrophysics $|$ Harvard \& Smithsonian, 60 Garden St., Cambridge, MA 02138, USA}

\author{Sang-Mok Cha}
\affiliation{Korea Astronomy and Space Science Institute, 776 Daedeokdae-ro, Yuseong-Gu, Daejeon 34055, Republic of Korea}
\affiliation{School of Space Research, Kyung Hee University, Giheung-gu, Yongin, Gyeonggi-do, 17104, Korea}

\author{Dong-Jin Kim}
\affiliation{Korea Astronomy and Space Science Institute, 776 Daedeokdae-ro, Yuseong-Gu, Daejeon 34055, Republic of Korea}

\author{Seung-Lee Kim}
\affiliation{Korea Astronomy and Space Science Institute, 776 Daedeokdae-ro, Yuseong-Gu, Daejeon 34055, Republic of Korea}

\author{Chung-Uk Lee}
\affiliation{Korea Astronomy and Space Science Institute, 776 Daedeokdae-ro, Yuseong-Gu, Daejeon 34055, Republic of Korea}

\author{Dong-Joo Lee}
\affiliation{Korea Astronomy and Space Science Institute, 776 Daedeokdae-ro, Yuseong-Gu, Daejeon 34055, Republic of Korea}

\author{Yongseok Lee}
\affiliation{Korea Astronomy and Space Science Institute, 776 Daedeokdae-ro, Yuseong-Gu, Daejeon 34055, Republic of Korea}
\affiliation{School of Space Research, Kyung Hee University, Giheung-gu, Yongin, Kyeonggi 17104, Republic of Korea}

\author{Byeong-Gon Park}
\affiliation{Korea Astronomy and Space Science Institute, 776 Daedeokdae-ro, Yuseong-Gu, Daejeon 34055, Republic of Korea}
\affiliation{University of Science and Technology, Korea, (UST), 217 Gajeong-ro, Yuseong-gu, Daejeon 34113, Republic of Korea}

\author{Richard W. Pogge}
\affiliation{Department of Astronomy, Ohio State University, 140 W. 18th Avenue, Columbus, OH 43210, USA}
\collaboration{(The KMTNet collaboration)}

\author{Radek Poleski}
\affiliation{Astronomical Observatory, University of Warsaw, AI.~Ujazdowskie~4, 00-478~Warszawa, Poland}

\author{Przemek Mróz}
\affiliation{Astronomical Observatory, University of Warsaw, AI.~Ujazdowskie~4, 00-478~Warszawa, Poland}

\author{Jan Skowron}
\affiliation{Astronomical Observatory, University of Warsaw, AI.~Ujazdowskie~4, 00-478~Warszawa, Poland}

\author{Michał K. Szymański}
\affiliation{Astronomical Observatory, University of Warsaw, AI.~Ujazdowskie~4, 00-478~Warszawa, Poland}

\author{Igor Soszyński}
\affiliation{Astronomical Observatory, University of Warsaw, AI.~Ujazdowskie~4, 00-478~Warszawa, Poland}

\author{Paweł Pietrukowicz}
\affiliation{Astronomical Observatory, University of Warsaw, AI.~Ujazdowskie~4, 00-478~Warszawa, Poland}

\author{Szymon Kozłowski}
\affiliation{Astronomical Observatory, University of Warsaw, AI.~Ujazdowskie~4, 00-478~Warszawa, Poland}

\author{Krzysztof Ulaczyk}
\affiliation{Astronomical Observatory, University of Warsaw, AI.~Ujazdowskie~4, 00-478~Warszawa, Poland}
\affiliation{Department of Physics, University of Warwick, Gibbet Hill Road, Coventry CV4 7AL, UK}

\author{Krzysztof A. Rybicki}
\affiliation{Astronomical Observatory, University of Warsaw, AI.~Ujazdowskie~4, 00-478~Warszawa, Poland}
\affiliation{Department of Particle Physics and Astrophysics, Weizmann Institute of Science, Rehovot 76100, Israel}

\author{Patryk Iwanek}
\affiliation{Astronomical Observatory, University of Warsaw, AI.~Ujazdowskie~4, 00-478~Warszawa, Poland}

\author{Marcin Wrona}
\affiliation{Astronomical Observatory, University of Warsaw, AI.~Ujazdowskie~4, 00-478~Warszawa, Poland}

\author{Mariusz Gromadzki}
\affiliation{Astronomical Observatory, University of Warsaw, AI.~Ujazdowskie~4, 00-478~Warszawa, Poland}
\collaboration{(The OGLE collaboration)}

\begin{abstract}
We report on the discovery and analysis of the planetary microlensing event \thisevent~ with a planet-to-star mass ratio $q \sim 0.003$.
The event \thisevent~ has unambiguous cusp-passing and caustic-crossing anomalies, which were caused by a wide planetary caustic with $s \simeq 2$, where $s$ is the star-planet separation in units of the angular Einstein radius $\thetae$.
Thanks to well-covered anomalies by the Korea Micorolensing Telescope Network (KMTNet), we measure both the angular Einstein radius and the microlens parallax in spite of a relatively short event timescale of $\te = 28\, \rm days$.
However,  because of a weak constraint on the parallax, we conduct a Bayesian analysis to estimate the physical lens parameters.
We find that the lens system is a super-Jupiter-mass planet of $M_{\rm p} = 1.75^{+0.53}_{-0.51}\, M_{\rm J}$ orbiting a late-type star of $M_{\rm h}=0.55^{+0.27}_{-0.26}\, M_\odot$ at a distance $\dl = 6.1^{+0.9}_{-1.3}\, \rm kpc$.
The projected star-planet separation is $a_{\perp} = 5.19^{+0.90}_{-1.23}\, \rm au$, which means that the planet orbits at about four times the snow line of the host star.
Considering the relative lens-source proper motion of $\murel = 6\, \rm mas\, yr^{-1}$, the lens will be separated from the source by $60\, \rm mas$ in 2029.
At that time one can measure the lens flux from adaptive optics imaging of Keck or a next-generation 30 m class telescope.
\thisevent Lb represents a growing population of wide-orbit planets detected by KMTNet, so we also present a general investigation into prospects for further expanding the sample of such planets.

\end{abstract}


\section{Introduction} \label{sec:intro}

As of 2023  April 10, 5332\footnote{https://exoplanetarchive.ipac.caltech.edu/index.html} exoplanets have been detected by various detection methods including transit, radial velocity, microlensing, and imaging.
Out of them, 95\% have been detected by transit and radial velocity methods, in which their host stars are mostly Sun-like stars and they are almost all located inside the snow line of the stars, which represents the distance where the water can form ice grains in the protoplanetary disk \citep{kennedy&kenyon2008}.
This is because the two methods depend on the brightness of stars and are advantageous in detecting close-in planets of the stars.
By contrast, microlensing is sensitive to planets around faint low-mass objects such as M dwarfs and brown dwarfs.
This is because the microlensing relies on the mass of objects, not their brightness.
As a result, microlensing exoplanets are almost all located beyond the snow line of their host stars and about 70\%\footnote{NASA Exoplanet Archive accessed 2023 April 10.} of the host stars are faint M dwarf stars.
Recently, \citet{shin+2023a} showed that M dwarf host stars with $M < 0.3\, M_\odot$ detected by microlensing commonly have massive planets with $> 0.3\, M_{\rm J}$, whereas for transit and radial velocity such massive planets are sparse in spite of high detection efficiency (see Figure 12 of \citet{shin+2023a}), which suggests that the formation of massive planets around low-mass M dwarfs is not challenging.
Therefore microlensing samples are very crucial to constrain planet formation theories, including core accretion and disk instability, which were constructed based on our solar system and modified by observed exoplanetary systems.
In addition, microlensing exoplanets are distributed in a wide range of distances from the Sun, $\sim 0.4 -8.8\, \rm kpc$, thus making it possible to investigate the census of all kinds of planets in the Galaxy.

The Korea Microlensing Telescope Netowrk (KMTNet; \citet{kim+2016}) has detected 124 (66\%) of the $187^{\,2}$ microlensing exoplanets discovered so far.
These results were achieved within about 7 yr after the start of the official observations of KMTNet in 2016, whereas the other 34\% of detections were achieved over about 25 yr.
This is due to 24 hr, high-cadence observations for a wide field of about 100 deg$^2$ toward the Galactic bulge \citep{shin+2016}.
\citet{chung+2022} reported that KMTNet has detected about 50\% of microlensing planets, as of 2022 April.
Compared with this, the planet discovery rate of KMTNet has increased by 16\% in a year.
This is mainly because the results of KMTNet's systematic AnomalyFinder (AF; \citealt{zang+2021}) applied to the 2018-2019 prime fields (\citealt{gould+2022, zang+2022}) and 2018 subprime fields \citep{jung+2022} have been published and they reported that 18 exoplanets have been newly discovered by the AF, which was initiated in order to find buried planetary signals.
Since the systematic AF searches of 2019 subprime \citep{jung+2023} and 2016 prime fields \citep{shin+2023b} were already done and other remaining seasons will be conducted soon, we expect that the KMTNet planet discovery rate will increase at a similar rate for the next few years.
In addition, thanks to the 24 hr high-cadence observations, KMTNet is readily detecting very-low-mass-ratio events of $q<10^{-4}$ by eye and via AF \citep{jung+2023, zang+2023},  which were rarely detected before KMTNet observations.
Moreover, KMTNet is often detecting planetary events caused by planetary caustics, which are more difficult to detect because planetary caustic anomalies are unpredictable.

In this paper, we analyze the microlensing event OGLE-2019-BLG-1180 and report a newly discovered wide-orbit giant planet around an M dwarf with KMTNet.
The paper is organized as follows.
In Section 2, we present the observations of \thisevent, and we describe the analysis of the light curve in Section 3.
We characterize the source star from its color and magnitude in Section 4, and we estimate the physical lens parameters from a Bayesian analysis in Section 5.
In Section 6, we review wide-orbit planetary lensing events and present a general investigation into prospects for further expanding the sample of wide-orbit planets.
Finally, we conclude in Section 7.

\section{Observations} \label{sec:obs}
The microlensing event OGLE-2019-BLG-1180 occurred at equatorial coordinates $({\rm RA, decl.})=(17:56:59.17,-27:58:31.6)$, corresponding to the Galactic coordinates $(l, b) = (2 \fdg 10, -1 \fdg 66)$.
The event was first alerted by the Optical Gravitational Lensing Experiment (OGLE; \citealt{udalski2003}).
OGLE uses a 1.3 m telescope with a $1.4\ \rm deg^{2}$ field of view (FOV) at Las Campanas Observatory in Chile.
The event lies in the OGLE IV field BLG504, which is observed with a cadence of $\Gamma \simeq 0.4\ \rm hr^{-1}$.
The OGLE data partially covered the anomalies induced by the wide planetary caustic, but it was not enough to find the best solution.
In addition, the peak was almost not covered by OGLE.

KMTNet also detected this event, and it was designated as KMT-2019-BLG-1912.
KMTNet uses 1.6 m telescopes with $4\ \rm deg^{2}$ FOV cameras at three different southern sites: the Cerro Tololo Inter-American Observatory in Chile (KMTC), the South African Astronomical Observatory (KMTS), and the Siding Spring Observatory in Australia (KMTA).
The event lies in the KMT main field BLG02 with a cadence of $\Gamma \simeq 2\, \rm hr^{-1}$.
With this high cadence, the peak and anomalies were well covered.
For the measurement of the color of the source star, KMTNet data were mainly taken in the $I$ band, and some data were taken in the $V$ band.
The KMTNet data were reduced using pySIS based on the difference imaging method (\citealt{alard&lupton1998}; \citealt{albrow+2009}).
For the characterization of the source color and construction of the color-magnitude diagram (CMD) of stars around the source, KMTC $I$- and $V$-band images were reduced using the pyDIA code \citep{albrow2017}.
However, all KMT $V$-band images for the three sites were affected by bleeding.
We thus could not construct the KMTC CMD of the event.
On the other hand, for the KMTC $I$-band images, the bleeding was located away from the source by 3 pixels ($1.2\arcsec$), but it was weak so that there was no problem for identifying the source.
For the $I$-band images of KMTA and KMTS, the bleeding was located away from the source by 7 pixels ($2.8\arcsec$), and thus the source was isolated enough.
Therefore, there was no problem to use the KMT $I$-band data for modeling.
For the source color and CMD, we used the OGLE $I$- and $V$-band data sets, which will be discussed in detail in Section 4.
The OGLE data were reduced by the difference imaging pipeline developed by \citet{wozniak2000}.

\section{Light-curve Analysis}
\subsection{Standard model}
The light curve of the event OGLE-2019-BLG-1180/KMT-2019-BLG-1912 has remarkable anomalies induced by cusp passing and caustic crossing.
We thus carry out standard binary lens modeling.
The binary lens modeling requires seven parameters: three single lensing parameters $(\tzero,\uo,\te)$, three binary lensing parameters $(s,q,\alpha)$, and the normalized source radius $\rho = \thetas/\thetae$, where $\thetas$ is the angular radius of the source star.
Here, $\tzero$ is the peak time of the event, $\uo$ is the lens-source separation in units of $\thetae$ at $t=\tzero$, $\te$ is the crossing time of $\thetae$, and $\alpha$ is the angle between the source trajectory and the binary axis.
In addition, there are two flux parameters $(\fss, \fb)$ for each observatory, which are the source flux and blended flux, respectively.
The two flux parameters $(\fss,\fb)$ are modeled  by $F(t) = \fss A(t) + \fb$, where $A(t)$ is the magnification as a function of time \citep{rhie+1999}, and they are determined from a linear fit.

The event has caustic-crossing anomalies, and thus we consider the limb-darkening variation of the finite source star in the modeling.
For this, we adopt the brightness variation of the source star, which is approximated by $S \propto 1 - \Gamma(1 - 3\cos\phi/2)$, where $\Gamma$ is the limb-darkening coefficient and $\phi$ is the angle between the normal to the source surface and the line of sight \citep{an+2002}.
According to the source type, which will be discussed in Section 4, we adopt $\Gamma_{I} = 0.43$ from \citet{claret2000}.

We first conduct a grid search in the binary lensing parameter space $(s, q, \alpha)$ to find local $\chi^2$ minima using the Markov Chain Monte Carlo (MCMC) method.
The ranges of each parameter are $-1 \leqslant \log s \leqslant 1$, $-4 \leqslant \log q \leqslant 0$, and $0 \leqslant \alpha \leqslant 2\pi$ with $(50, 50, 20)$ uniform grid steps, respectively.
In the grid search, $s$ and $q$ are fixed, while the other parameters are allowed to vary in the MCMC chain.
From the grid search, we find three local solutions $(s,q)=(0.543,0.106)$, $(s,q)=(1.677,0.002)$, and $(s,q)=(2.024,0.017)$.
We then carry out an additional modeling in which the local solutions are set to the initial values and all parameters are allowed to vary.
As a result, we find that the two wide models converge to $(s,q)=(1.89,0.005)$ and the $\chi^2$ of the close model is much larger than that of the wide model by $926$, and thus the event was caused by a wide planetary system.
However, the best-fit light curve of the wide planetary system does not fit well to the data, especially at the regions with anomalies.
This indicates that the event would be likely affected by high-order effects including the microlens parallax and lens orbital motion effects.

\begin{figure}[t!]
\centering
\includegraphics[width=0.5\textwidth]{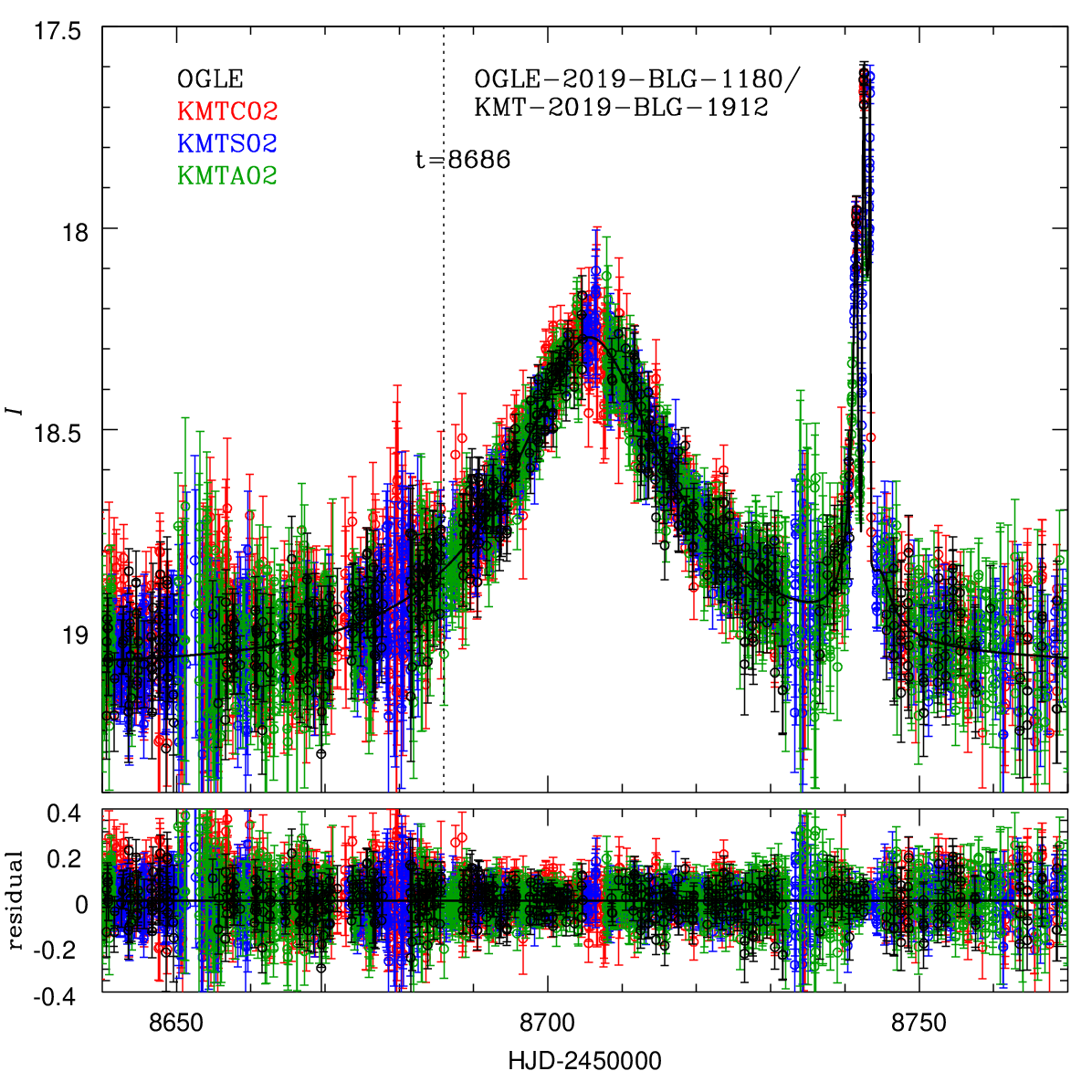}
\caption{Light curve of the best-fit parallax+orbital lens model of \thisevent.\\
(The data used to create this figure are available)
\label{fig:best}}
\end{figure}

\subsection{High-order effects}
\subsubsection{Parallax+Orbital model}

\begin{figure*}[t!]
\centering
\includegraphics[width=0.9\textwidth]{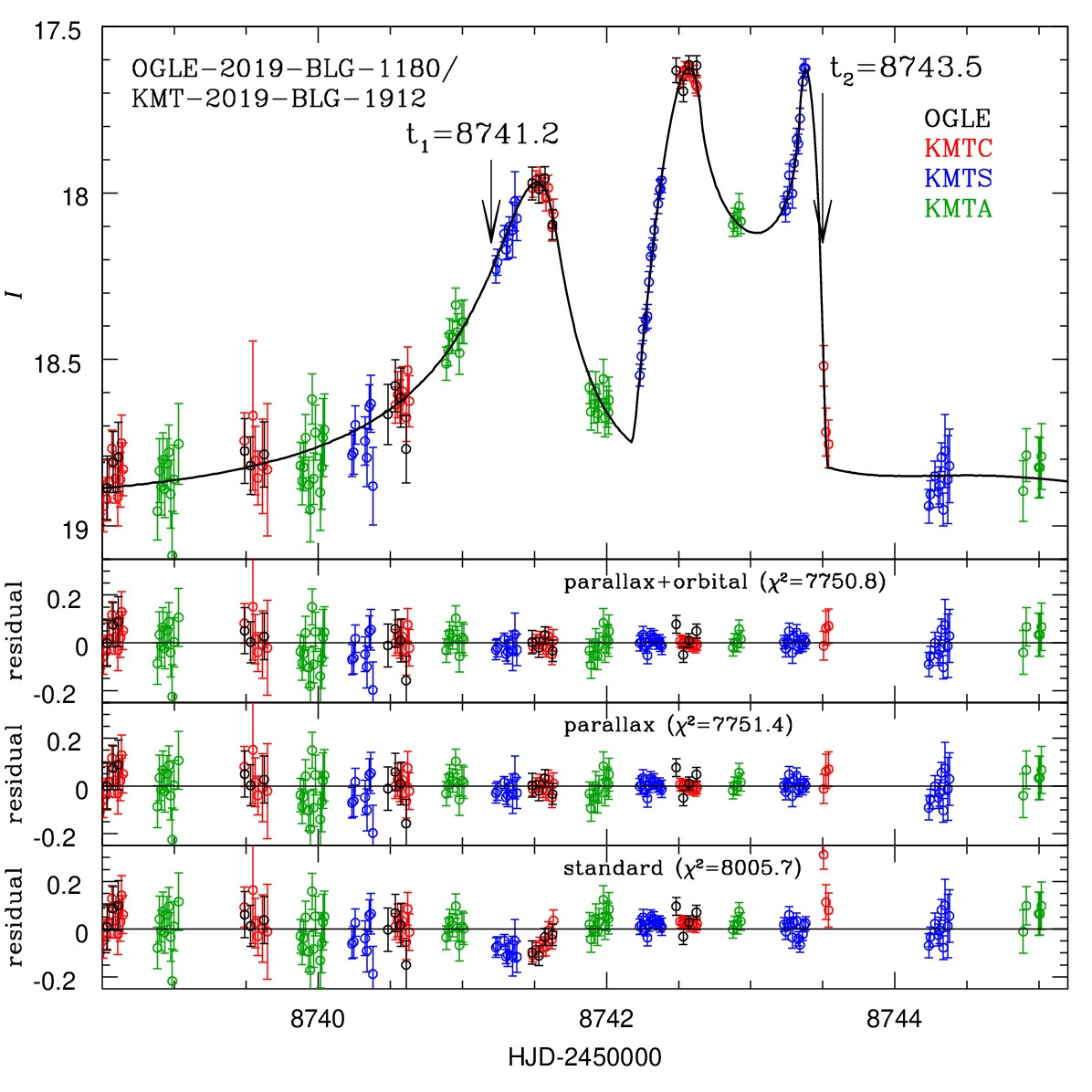}
\caption{Close-up view of the anomaly region.
Here $t_1$ and $t_2$ represent the times before the source passes the cusp and right after the source exits the caustic, respectively (see Figure \ref{fig:geom}).
\label{fig:closeup}}
\end{figure*}

\begin{figure}[t!]
\centering
\includegraphics[width=0.5\textwidth]{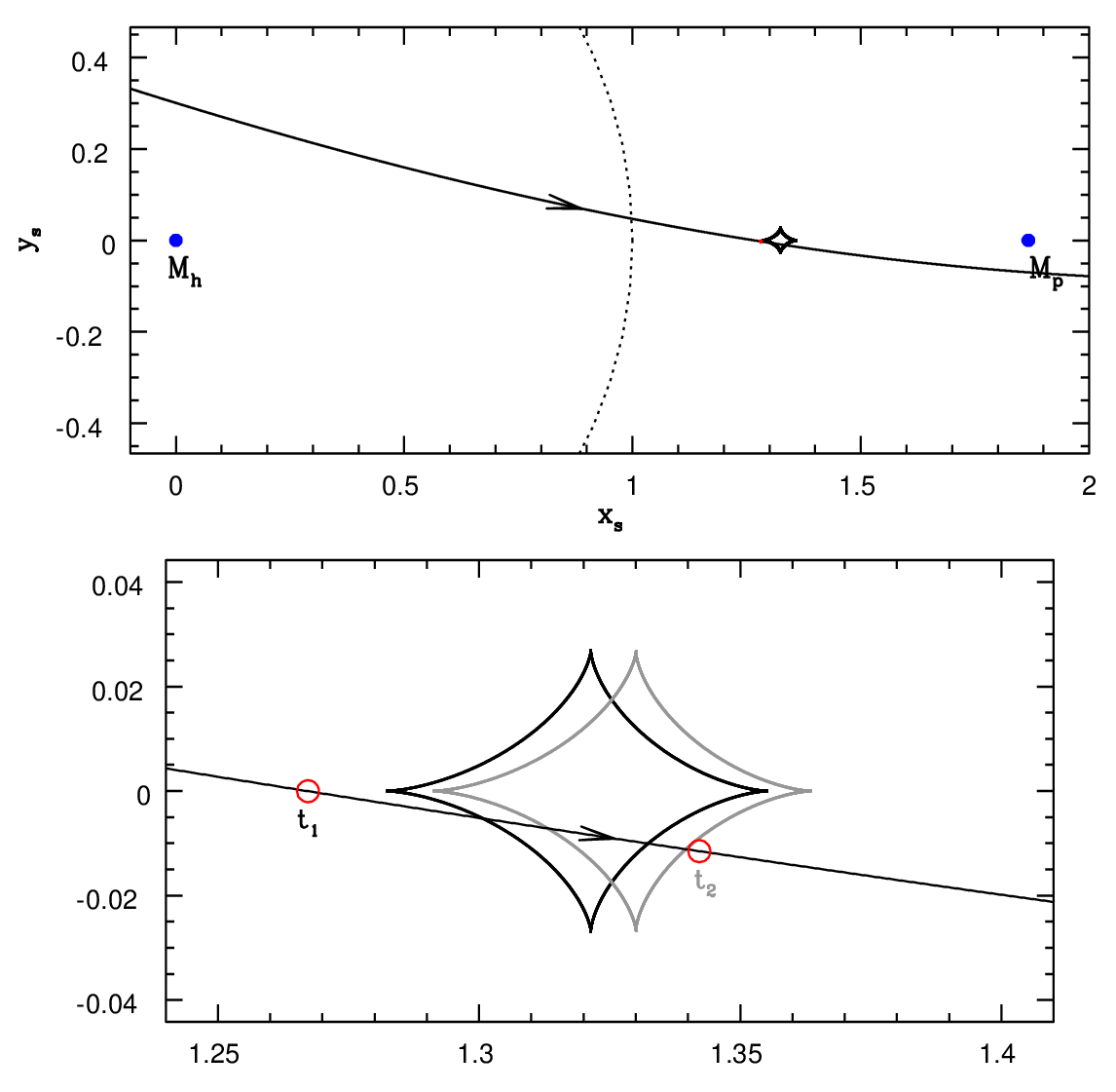}
\caption{Geometry of the best-fit parallax+orbital model.
Top:the blue solid dots represent two lens components, while the red open circle represents the normalized source size.
The dotted circle denotes the Einstein ring and the straight line with an arrow is the source trajectory.
The black closed curve represents the planetary caustic. 
Bottom: close-up view of the planetary caustic region.
The caustics at $t_{1}=8741.2$ and $t_{2}=8743.5$ are presented in black and gray, respectively.
\label{fig:geom}}
\end{figure}

\begin{figure}[t!]
\centering
\includegraphics[width=0.5\textwidth]{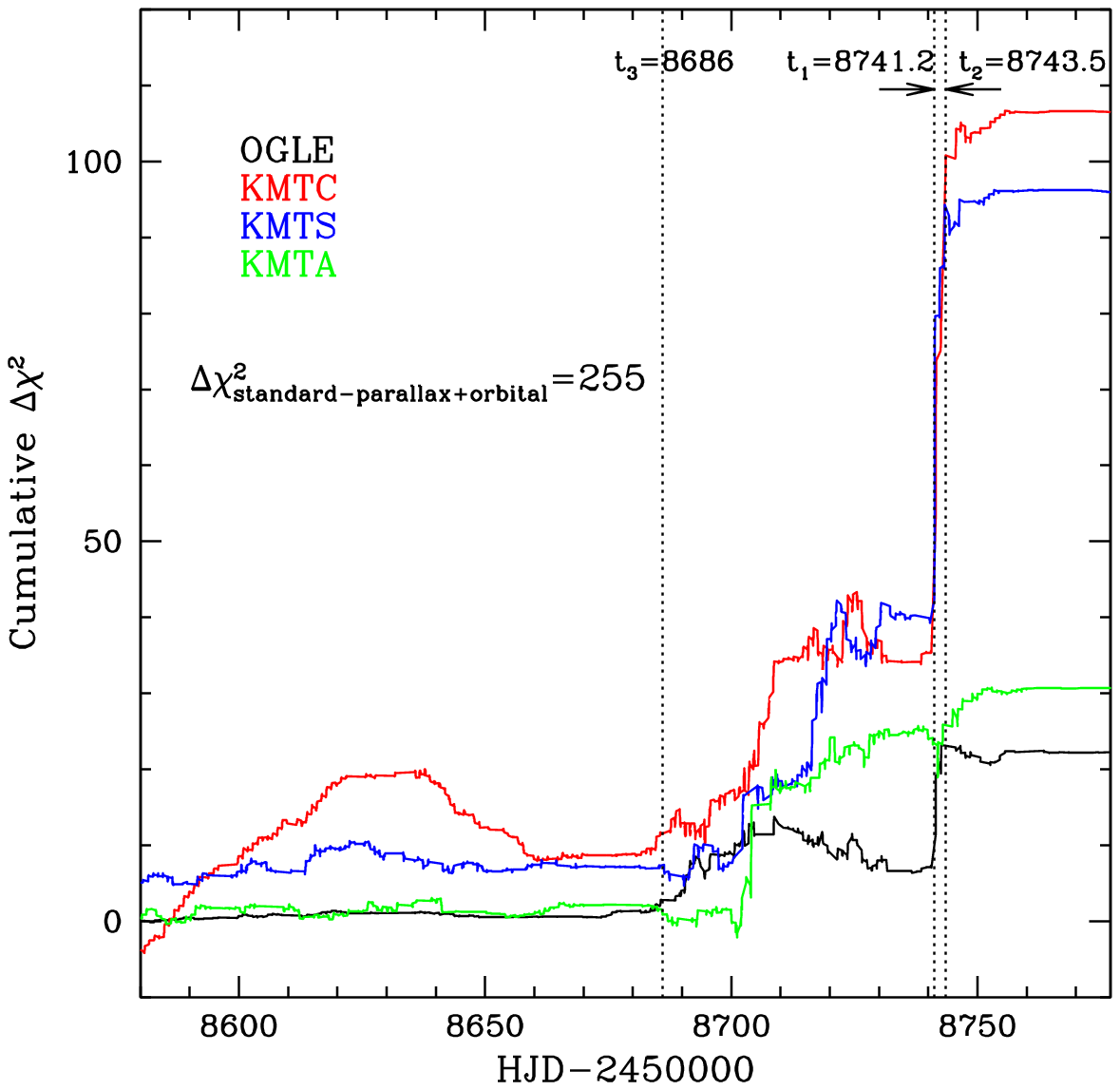}
\caption{Cumulative $\delcs$ between the standard and the parallax+orbital models.
\label{fig:cum}}
\end{figure}

The microlens parallax is usually well measured for events with a long timescale ($\te > \rm{yr}/2\pi$; \citealt{yoo+2004}), because it is caused by the orbital motion of the Earth.
The timescale of this event is $\te=28\, \rm days$, which is relatively short to be affected by the microlens parallax.
However, the event has both cusp-passing and caustic-crossing anomalies induced by the wide planetary caustic and the anomalies were well covered.
In this case, it is possible to measure the microlens parallax.
Since the lens orbital motion effect can mimic the parallax effect (\citealt{batista+2011}; \citealt{skowron+2011}), we conduct the parallax modeling together with the orbital motion effect.
The microlens parallax is described by $\bpie=(\pien, \piee)$, while the lens orbital motion is described by $(ds/dt, d\alpha/dt)$, which are the instantaneous changes of the binary separation and the orientation angle of the binary axis, respectively.
Here we note that the orbital parameters are not well constrained, and thus we consider only lens systems with a ratio of the projected kinetic to the potential energy limited to $\beta < 0.8$.
The ratio $\beta$ is defined as
\begin{equation}
\beta \equiv \left({\rm KE\over{\rm PE}}\right)_\perp = {{(s\thetae D_{\rm L}/{\rm au}})^3(\gamma^2\ {\rm yr}^2)\over{M/M_\odot}},
\end{equation}
where $\gamma = [(ds/dt/s)^2 + (d\alpha/dt)^2]^{1/2}$.
As a result, we find that the whole data set is well fitted by the parallax+orbital model, especially, in the region with anomalies, and its $\chi^2$ is improved by $255$ compared to the standard model.
The best-fit light curve of the parallax+orbital model is presented in Figure \ref{fig:best}, and Figure \ref{fig:closeup} shows a close-up view of the region with anomalies.
The best-fit lensing parameters are presented in Table \ref{tab-best} and the geometry of the best-fit model is shown in Figure \ref{fig:geom}. 

\begin{figure}[t!]
\centering
\includegraphics[width=0.5\textwidth]{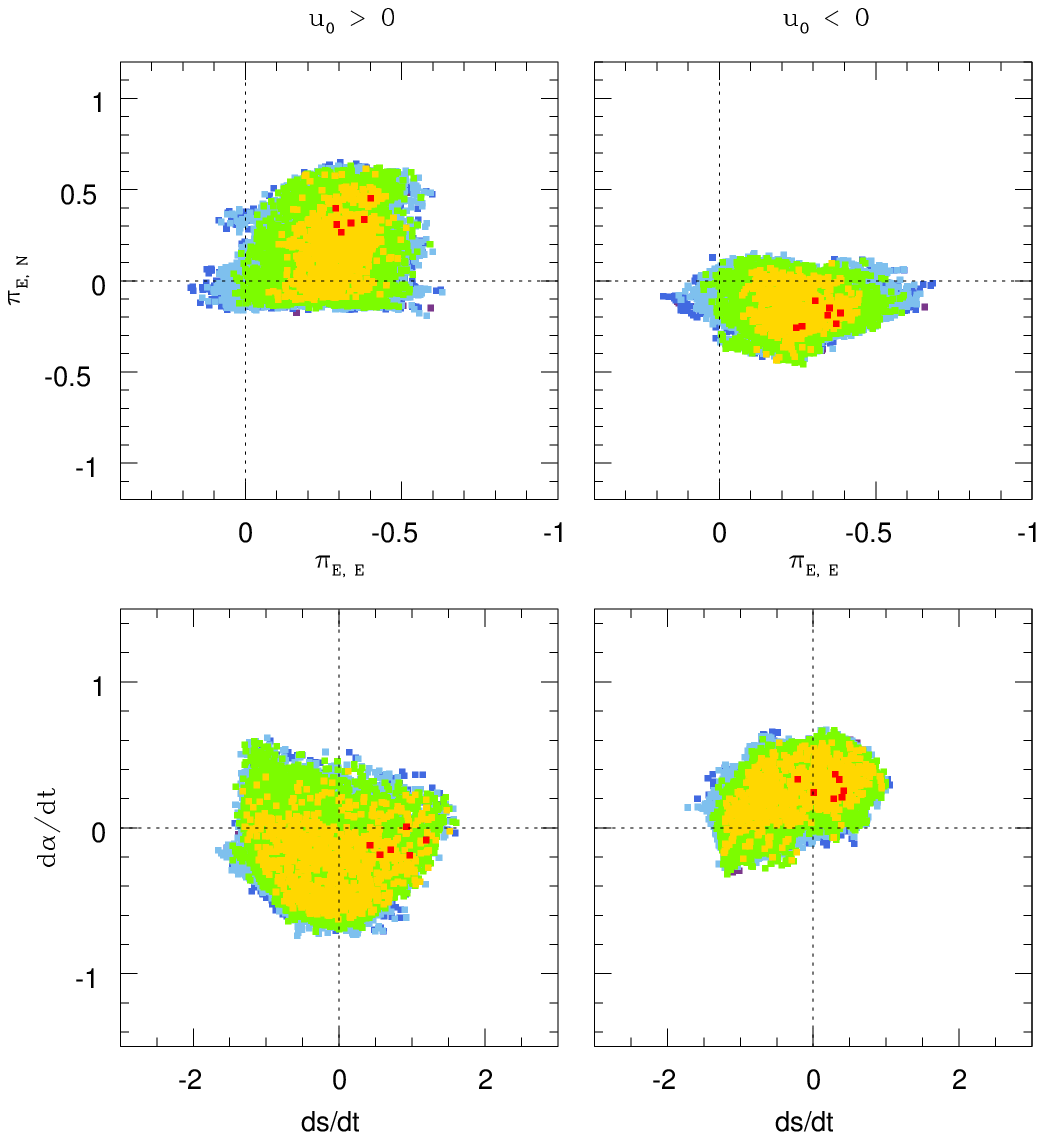}
\caption{$\chi^2$ distributions of the best-fit parallax+orbital model with the partial data sets.
The red, yellow, green, light blue, dark blue, and purple represent regions with $\delcs < (1, 4, 9, 16, 25, 36)$, from the best-fit model, respectively.
\label{fig:p+o}}
\end{figure}

In order to find the source of the $\chi^2$ improvement, we construct the cumulative distribution of $\delcs$ between the standard and the parallax+orbital models as a function of time.
As shown in Figure \ref{fig:cum}, only the KMTC and KMTS data have improvements in the range of HJD$^\prime = {\rm HJD}-2450000 < 8686$, while for OGLE and KMTA, there is no improvement.
However, we could not find noticeable improvements of the light curve in the range of HJD$^\prime < 8686$ for KMTC and KMTS.
This trend is the same as the event OGLE-2018-BLG-1428 \citep{kim+2021}.
As mentioned in \citet{kim+2021}, the improvements of KMTC and KMTS are likely due to correlated noise.
On the other hand, for the range of HJD$^\prime \geqslant 8686$, the improvement of KMTC and KMTS looks reasonable.
This is because the improvement of KMTC follows OGLE at the same time zone, but having better improvements due to better coverage (especially the anomaly range) relative to OGLE, while for KMTS it is similar to KMTC due to the similar coverage.
As shown in Figure \ref{fig:closeup}, the anomalies were almost covered by KMTC and KMTS.
Hence,  we carry out parallax+orbital remodeling with partial data sets for KMTC and KMTS and full data sets for OGLE and KMTA.
For KMTC and KMTS, we use only the data sets in the range HJD$^\prime \geq 8686$.
Figure \ref{fig:p+o} shows the $\chi^2$ distributions of the best-fit parallax and orbital parameters.
From Figure \ref{fig:p+o}, we find that the parallax ($\pie=0.46 \pm 0.14$) has relatively large errors, while the orbital parameters were very poorly constrained.
The best-fit lensing parameters for the partial data sets are presented in Table \ref{tab-best-partial}.

\subsubsection{Parallax-only model}
\begin{figure}[t!]
\centering
\includegraphics[width=0.5\textwidth]{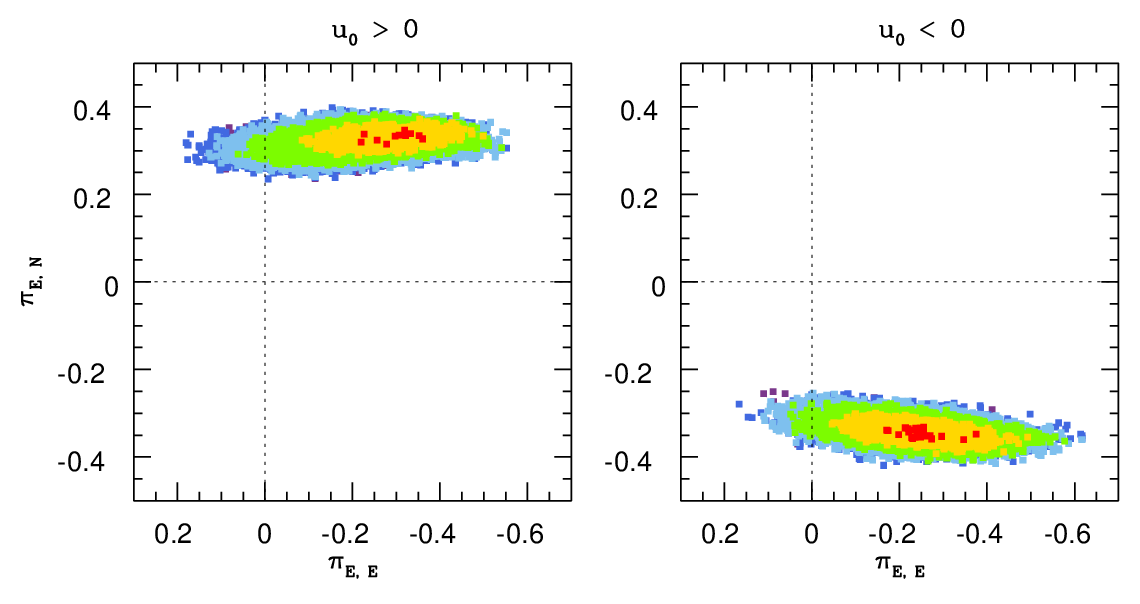}
\caption{$\chi^{2}$ distributions of the parallax-only model for the partial data sets.
\label{fig:parallax}}
\end{figure}

\begin{figure}[t!]
\centering
\includegraphics[width=0.5\textwidth]{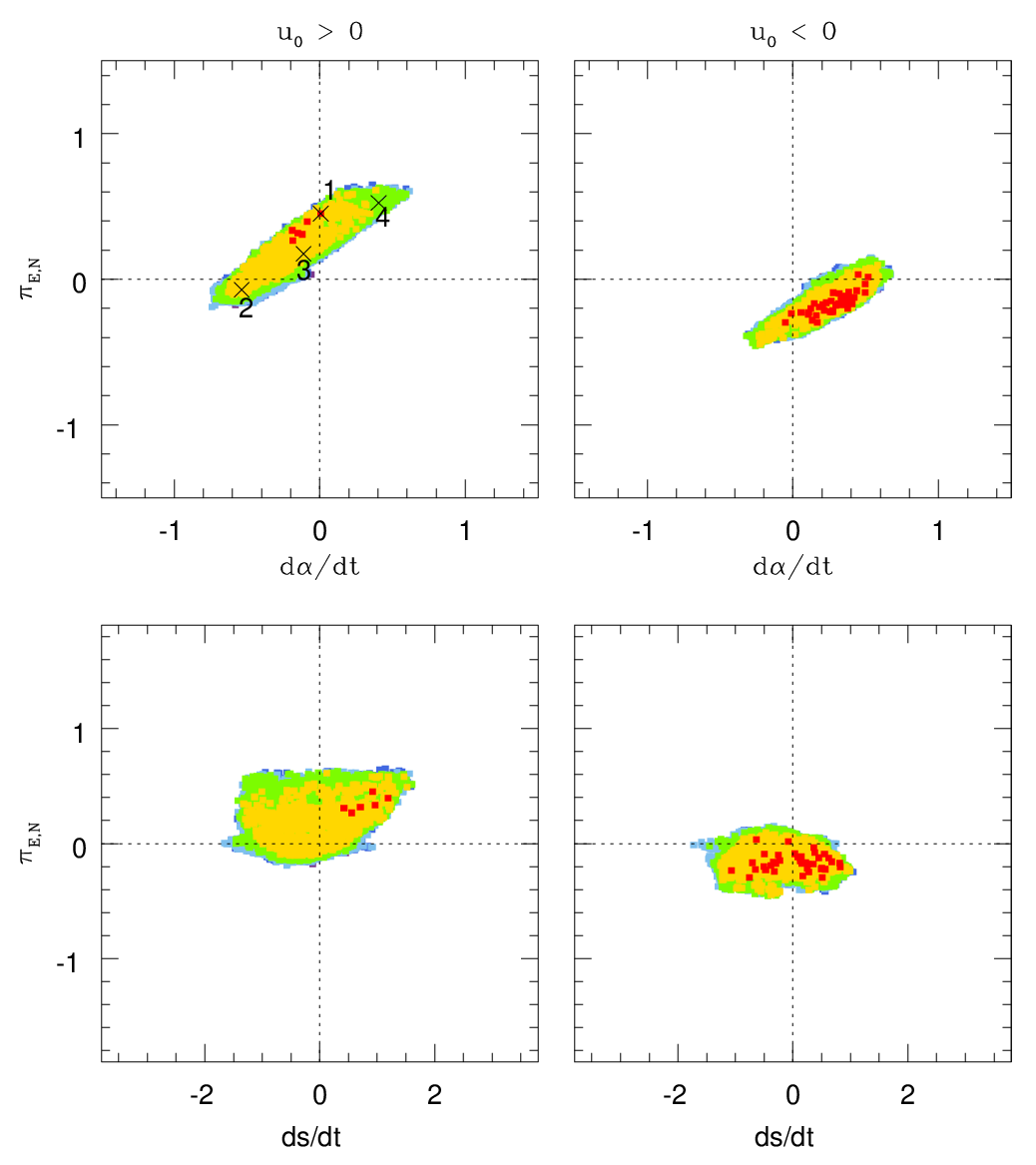}
\caption{$\chi^2$ distributions of the parallax and orbital parameters.
Four models with $\delcs < 4$ are marked as 1, 2, 3, and 4 in the distribution of $u_{0} > 0$; their corresponding parameters are presented in Table \ref{tab-degen}.\label{fig:degen}}
\end{figure}

In order to check why the parallax and orbital parameters are not well constrained, in spite of the well-covered anomalies, we conduct parallax-only modeling.
The resulting best-fit parameters for the full data sets and partial data sets are presented in Tables \ref{tab-best} and \ref{tab-best-partial}, respectively.
Figure \ref{fig:parallax} shows the $\chi^2$ distribution of the best-fit parallax-only model for the partial data sets.
As shown in the figure, we find that $\pien$ is much better constrained relative to the parallax+orbital model and the $\delcs$ between the parallax-only and parallax+orbital models is very small by $\delcs = 0.6$, while $\piee$ is similarly constrained to that of the parallax+orbital model.
The big improvement of the $\pien$ error indicates that the parallax parameters are strongly correlated with the orbital parameters.
We plot the $\chi^2$ of the parallax and orbital parameters to check their correlation in Figure \ref{fig:degen}.
From Figure \ref{fig:degen}, we find a strong correlation between $\pien$ and $d\alpha/dt$, while $\pien$ is barely correlated with $ds/dt$.
Hence, even though the orbital parameters are not well constrained, it is clear that they are important because including them changes the parallax constraints; i.e., part of the reason the orbital parameters are poorly constrained is because of the degeneracy with the parallax parameters.
We also mark four different models in Figure \ref{fig:degen} and their parameters are presented in Table \ref{tab-degen}.
As shown in Table \ref{tab-degen}, each model has different parallax and orbital parameters, but the $\delcs$ among the four models is small ($\delcs < 4$) and the physical parameters of the models are similar.
This result is similar to the results of \citet{batista+2011} and \citet{skowron+2011}, where there is a strong degeneracy between $\pien$ and $d\alpha/dt$, and shows that it is very important to consider simultaneously the parallax and orbital parameters, even in the cases where the orbital parameters are poorly constrained.

\begin{deluxetable*}{lcccccccc}
\tablewidth{0pt}
\tablecaption{Lensing parameters of OGLE-2019-BLG-1180\label{tab-best}}
\tablehead{
                                              &    \multicolumn{2}{c}{Standard}          &&     \multicolumn{2}{c}{Parallax}               &&\multicolumn{2}{c}{Parallax+Orbital} \\
\cline{2-9}
Parameter                              &     $u_{0}>0$      &    $u_{0}<0$         &&   $u_{0}>0$     &   $u_{0}<0$                 && $u_{0}>0$       &    $u_{0}<0$          
 }
\startdata
$\chi^2$                                &   $8005.74$      &  $8005.94$         &&    $7751.45$     &  $7752.74$                   &&  $7750.80$       &  $7752.08$ \\
$t_0$ (HJD$^\prime$)        &   $8705.7077$   &  $8705.7005$      &&  $8705.7673$ &  $8705.7610$              &&  $8705.7619 $  &  $8705.7931$ \\
                                              &   $(0.0420)$      &  $(0.0420)$          &&   $(0.0580)$      &  $(0.0582)$                 &&   $(0.0648)$     &  $(0.0612)$ \\
$u_0$                                   &  $0.2614$         &  $-0.2612$           &&  $0.2880$         &  $-0.2891$                   &&  $0.2876$        &  $-0.2876$ \\
                                              &   $(0.0013)$       &  $(0.0013)$         &&   $(0.0024)$    &  $(0.0025)$                  &&  $(0.0027)$      &  $(0.0027)$ \\
$\te$ (days)                         &  $27.8869$       &  $27.8715$          &&  $28.1782$       &  $28.0997$                   &&  $28.1627$       &  $28.0510$ \\
                                             &  $(0.1362)$        &  $(0.1389)$          &&   $(0.1779)$      &  $(0.1731)$                   &&  $(0.1902)$      &  $(0.2051)$ \\
$s$                                       &  $1.8936$         &  $1.8943$             &&   $1.8656$        &  $1.8645$                     &&  $1.8642$          &  $1.8706$ \\
                                             &  $(0.0051)$       &  $(0.0052)$         &&  $(0.0070)$       &  $(0.0075)$                  &&  $(0.0079)$       & $(0.0076)$ \\
$q\,(10^{-3})$                       &  $4.7969$         &   $4.8275$             &&  $3.5516$         &  $3.5075$                     &&  $2.8209$         &  $3.3405$  \\
                                             &  $(0.1139)$        &  $(0.1165)$         &&  $(0.2154)$       &  $(0.2233)$                  &&  $(0.5583)$      &  $(0.4776)$ \\
$\alpha$ (radians)                     &  $0.2021$        &  $-0.2018$            &&   $0.2772 $       &   $-0.2771$                   && $0.2818$          &  $-0.2414$ \\
                                             &  $(0.0013)$       &  $(0.0013)$          &&  $(0.0051)$       &  $(0.0049)$                  &&  $(0.0188)$       &  $(0.0183)$ \\
$\rho$                                  &  $0.0024$         &   $0.0024$           &&  $0.0023$        &   $0.0023$                     &&  $0.0021$         &  $0.0023$ \\
                                             &  $(0.0001)$       &  $(0.0001)$         &&  $(0.0001) $     &  $(0.0001)$                  &&  $(0.0002)$      &  $(0.0001)$\\
$\pien$                                 &     ...                    &    ...                         &&   $0.3616 $      &   $-0.3757$                 &&  $0.4076$         &  $-0.1208$ \\
                                             &     ...                     &    ...                         &&  $(0.0198)$     &  $(0.0201)$                 &&   $(0.1421)$     &  $(0.1411)$ \\
$\piee$                                 &     ...                    &    ...                         &&   $-0.2896$     &   $-0.2894$                 &&  $-0.2891$       &   $-0.1804$ \\
                                             &     ...                     &    ...                         &&  $(0.0874)$     &  $(0.0901)$               &&   $(0.0911)$     &  $(0.0921)$ \\
$ds/dt$ (yr$^{-1}$)            &     ...                    &    ...                         &&   ...                      &   ...                                &&  $1.0686$         &   $0.5397$ \\
                                             &     ...                     &    ...                        &&  ...                       &   ...                                &&   $(0.6235)$      &  $(0.5575)$ \\
$d\alpha/dt$ (radians yr$^{-1}$) &     ...                &   ...                         &&  ...                        &   ...                                &&  $-0.0902$       &  $0.4517$ \\
                                             &     ...                    &    ...                        &&  ...                        &   ...                                &&   $(0.2244)$     &  $(0.2216)$ \\
$f_{\rm s, ogle}$                 &  $0.1482$         &  $0.1483$            &&   $0.1612$         &   $0.1619$                   &&  $0.1610$          &  $0.1611$ \\
                                             &   $(0.0007)$     &  $(0.0007)$         &&   $(0.0013)$      &  $(0.0013)$                &&  $(0.0014)$       &  $(0.0014)$ \\
$f_{\rm b, ogle}$                &  $0.2194$         &  $0.2194$             &&   $0.2056$       &  $0.2050$                   &&  $0.2059$        &  $0.2058$ \\
                                            &   $(0.0007)$     &  $(0.0007)$          &&    $(0.0013)$    &  $(0.0013)$                 &&  $(0.0014)$      &  $(0.0014)$ \\
\enddata 
\tablecomments{ HJD$^\prime$ = HJD - 2450000.
The value in parantheses represents 1$\sigma$ error of each parameter.}
\end{deluxetable*}

\begin{deluxetable*}{lcccccc}
\tablewidth{0pt}
\tablecaption{Lensing parameters for the Parallax-only and Parallax+Orbital model with the Partial KMTC and KMTS data sets and the Full OGLE and KMTA data sets\label{tab-best-partial}}
\tablehead{
                                                                   & \multicolumn{2}{c}{Parallax}                                  &&                      \multicolumn{2}{c}{Parallax+Orbital}  \\
\cline{2-6}                                              
Parameter                          &         $u_{0}>0$             &           $u_{0}<0$                                     &&             $u_{0}>0$                 &           $u_{0}<0$              
 }
\startdata
$\chi^2$                                &        $4980.49$                       &   $4980.96$                              &&         $4979.85$                      &   $4980.20$                         \\
$t_0$ (HJD$^\prime$)         &  $8705.7512 \pm 0.0658$     &   $8705.7693 \pm 0.0620$     && $8705.7423 \pm 0.0676$      &   $8705.7945 \pm 0.0659$ \\
$u_0$                                    &  $0.2863 \pm 0.0030$          &   $-0.2868 \pm 0.0026$          &&  $0.2877 \pm 0.0032$           &   $-0.2862 \pm 0.0032$      \\
$\te$ (days)                           &  $27.7722 \pm 0.1938$         &   $27.7904 \pm 0.1870$            &&  $27.8479 \pm 0.2206$        &   $27.8238 \pm 0.2016$      \\
$s$                                          &  $1.8739 \pm 0.0103$          &   $1.8735 \pm 0.0089$             &&  $1.8673\pm 0.0100$            &   $1.8741 \pm 0.0107$        \\
$q\, (10^{-3})$                        &    $3.6282 \pm 0.3231$        &  $3.6693 \pm 0.2684$             &&  $3.3079 \pm 0.5456$          &  $3.4273 \pm 0.4966$          \\
$\alpha$ (radians)                        &   $0.2709 \pm 0.0070$         &   $-0.2698 \pm 0.0056$          &&  $0.2699 \pm 0.0221$           &   $-0.2567 \pm 0.0131$      \\
$\rho$                                     &   $0.0024 \pm 0.0001$          &  $0.0023 \pm 0.0001$            &&  $0.0021 \pm 0.0002$           &  $0.0023 \pm 0.0001$         \\
$\pien$                                    &   $0.3390 \pm 0.0237$         &   $-0.3528 \pm 0.0222$        && $0.3170 \pm 0.1641$              &   $-0.2572 \pm 0.0964$      \\
$\piee$                                    &    $-0.3331 \pm 0.1181$         &   $-0. 2965 \pm 0.1038$         &&  $-0.3378 \pm 0.1172$           &   $-0.2456 \pm 0.1225$       \\
$ds/dt$ (yr$^{-1}$)                &               ...                                &                ...                                 &&  $0.7098 \pm 0.5537$            &   $0.3941 \pm 0.5153$        \\
$d\alpha/dt$ (radians yr$^{-1}$) &               ...                                &                ...                                 &&  $-0.1498 \pm 0.2403$          &   $0.2086 \pm 0.1644$         \\
$f_{\rm s, ogle}$                     &     $0.1614 \pm 0.0015$        &   $0.1615 \pm 0.0014$             &&  $0.1615 \pm 0.0017$             &   $0.1610 \pm 0.0017$        \\
$f_{\rm b, ogle}$                     &     $0.2057 \pm 0.0015$       &  $0.2056 \pm 0.0014$            &&  $0.2056 \pm 0.0016$           &  $0.2061 \pm 0.0017$          \\
\enddata
\end{deluxetable*}

\subsubsection{Xallarap model}

\begin{figure}[t!]
\centering
\includegraphics[width=0.5\textwidth]{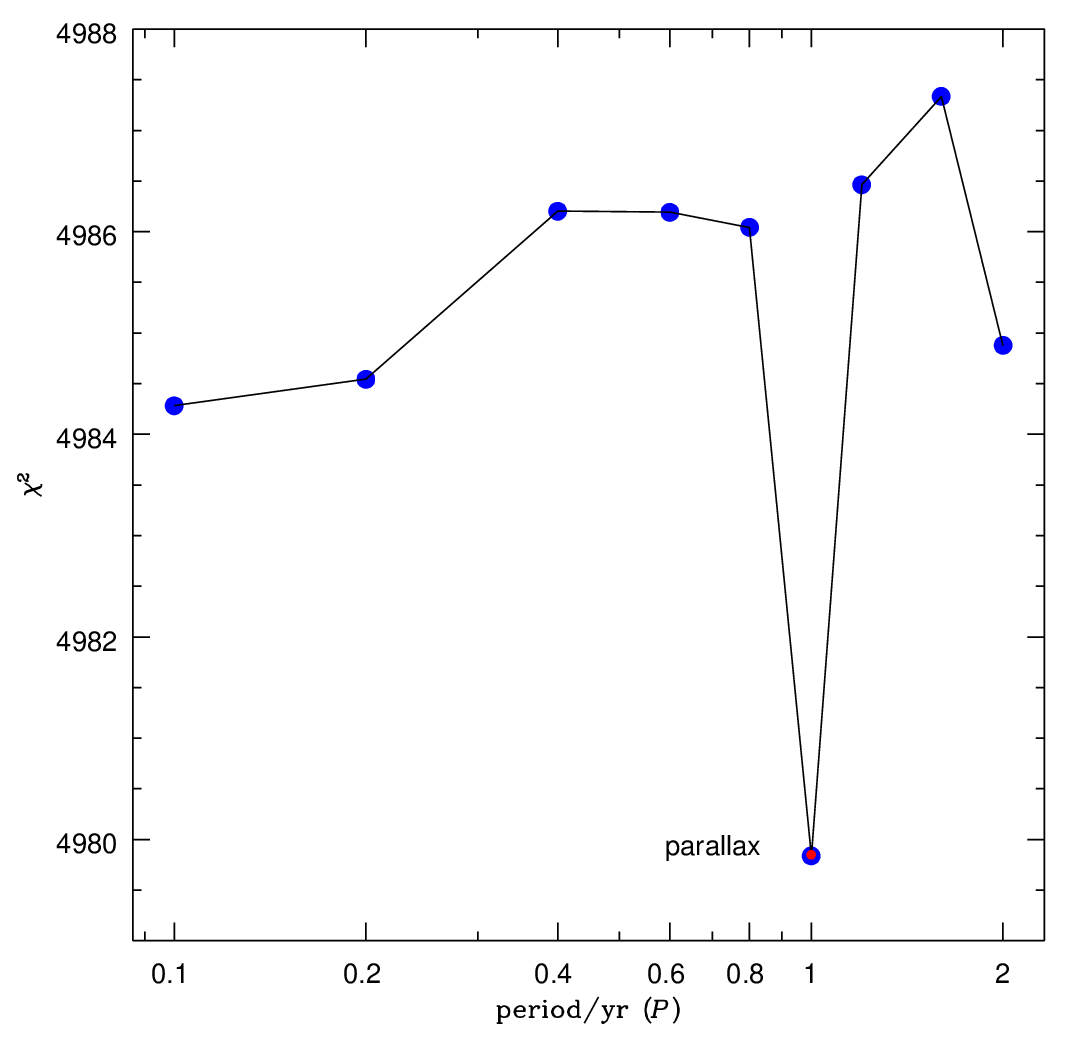}
\caption{$\chi^2$ distribution for the best-fit xallarap solutions as a function of fixed binary source orbital period $P$.
The red dot denotes the $\chi^2$ of the best-fit parallax model.
\label{fig:xallarap}}
\end{figure}

The parallax signal could come from the xallarap effect because of the orbital motion of the binary source.
We thus conduct xallarap modeling.
In the modeling, we assume that the binary source is in a circular orbit.
The xallarap modeling requires five additional parameters from the standard model: the orbital period of the binary source $P$, the counterparts of the parallax parameters ${\bm \xi_{\rm E}}=(\xi_{\rm E,N}, \xi_{\rm E,E})$, and the phase $\lambda$ and inclination $i$ of the binary source's orbital motion.
If the parallax measurement is real, the xallarap period should converge to the Earth orbital motion of 1 yr.
This is because the parallax effect is caused by the orbital motion of the Earth, as mentioned in Section 3.2.1.
Figure \ref{fig:xallarap} shows the $\chi^2$ distributions for the best-fit xallarap solutions as a function of fixed orbital period for the binary source $P$.
As shown in the figure, the best-fit xallarap solution appears at $P=1\, \rm yr$.
The $\chi^2$ difference between the xallarap and the parallax models is $\delcs = 0.5$, indicating that they are almost the same.
These suggest that the parallax measurement is real.

We then check that the best-fit xallarap solution is physically reasonable.
With $P$ and $(\xi_{\rm E,N}, \xi_{\rm E,E})$, we can check the reasonability of the xallarap solution.
$\xi_{\rm E}$ is defined as
\begin{equation}
\xi_{\rm E} = {a_{\rm s}\over{\hat{r}_{\rm E}}},\quad  {\hat{r}_{\rm E}\over{\rm au}} = \thetae D_{\rm S}
\end{equation}
where $a_s$ is the semi-major axis of the binary source, $\hat{r}_{\rm E}$ is the Einstein radius projected on the source plane, and $D_{\rm S}$ is the distance to the source.
We adopt $D_{\rm S}=7.73\ \rm kpc$ from \citet{nataf+2013}.
The source is a late G dwarf star in the bulge, and thus $\thetae = 0.46\,\rm mas$ (see Section 4).
Hence, the mass of the source is $\sim 1.0\, \rm M_\odot$ and $\hat{r}_{\rm E }=3.54\, \rm au$. 
We then apply Kepler's third law
\begin{equation}
{M_{\rm tot}\over{M_\odot}}\left(P\over{\rm yr}\right)^{2} = {\left(a_{\rm tot}\over{\rm au}\right)^{3}},
\end{equation}
where $M_{\rm tot} = M_{\rm s} + M_{\rm comp}$ and $a_{\rm s}/a_{\rm tot} = M_{\rm comp}/M_{\rm tot}$ (\citealt{dong+2009,kim+2021}).
Here $M_{\rm s}$ and $M_{\rm comp}$ are the masses of the source and its companion, respectively.
As mentioned in \citet{kim+2021}, Equation (3) can be described as a cubic equation
\begin{equation}
{(1+Q)^2\over{Q^3}} = {M_{\rm s}\over{M_\odot}} {(P/{\rm yr})^{2}\over{a_{\rm s}^{3}}}; \quad Q={M_{\rm comp}\over{M_{\rm s}}}.
\end{equation}
If  $0.1 < Q < 1.0$, the solution is physically reasonable, which means that the companion will be a typical main-sequence star.
For the best-fit xallarap solution, the orbital period of the source is $P=1\, \rm yr$ and $\xi_{\rm E}=0.47$, and thus $Q=6.1$.
We can also estimate the minimum mass of the source companion with $\theta_{\rm E,min}$, which is defined as $M_{\rm c,min}={(\xi_{\rm E}\hat{r}_{\rm E,min}/\rm au)^3/{(P/yr)^2}}$.  Adopting $\theta_{\rm E, min} \simeq 0.41\, \rm mas$ from Section 4, the minimum mass of the source companion is $M_{\rm c,min} \simeq 3.2\, M_\odot$.
These indicate that the companion would be a black hole, and thus it is not physically reasonable.
Therefore, we can rule out the xallarap interpretation.

\subsubsection{Point-source Point-lens model with Parallax}

\begin{figure}[t!]
\centering
\includegraphics[width=0.5\textwidth]{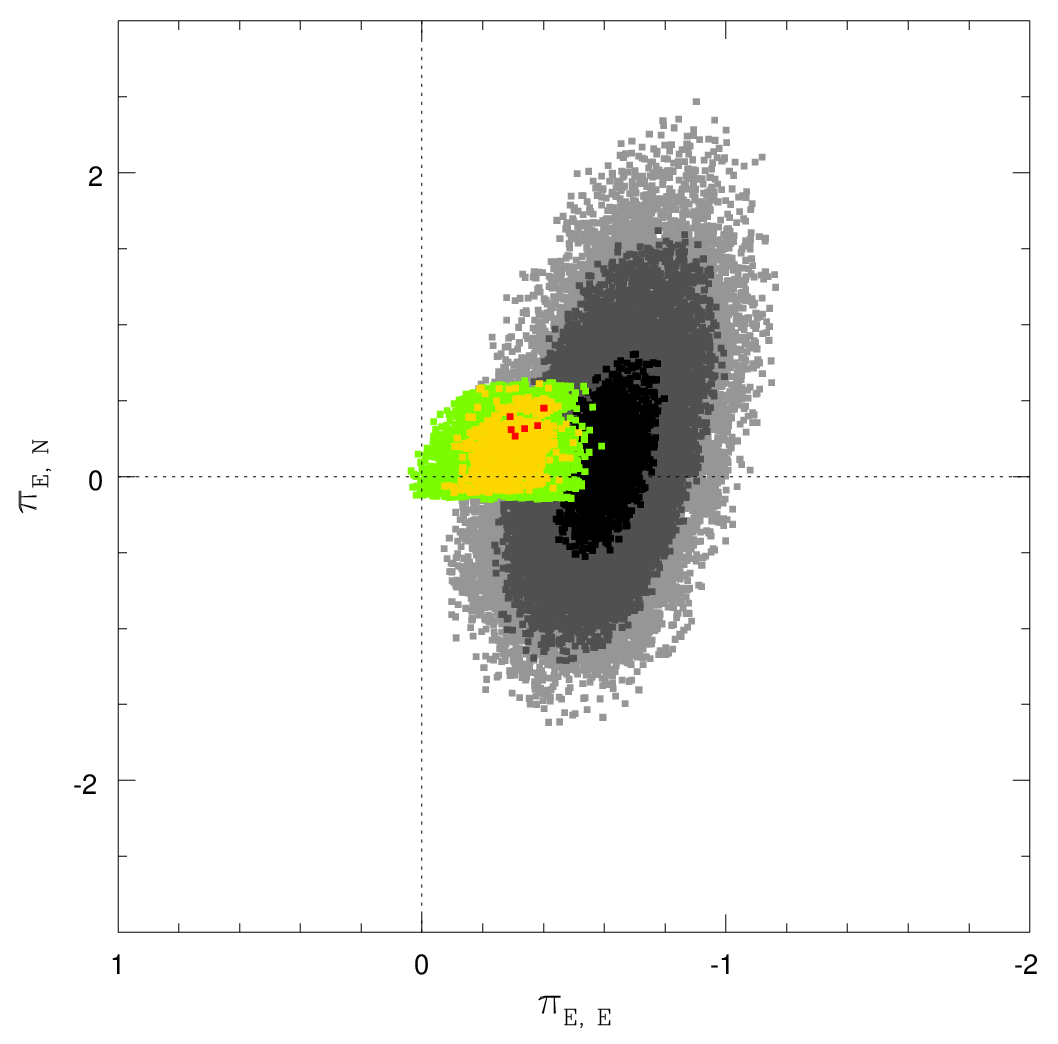}
\caption{$\chi^2$ distribution for the best-fit PSPL+parallax model (black and gray plots) together with the parallax+orbital model (colored plots).
We note that the PSPL+parallax modeling was conducted using the data sets that had the planetary anomaly region $8725 < {\rm HJD}^\prime < 8755$ removed.
\label{fig:pspl}}
\end{figure}

In order to test the impact of the planetary anomaly on the parallax, we conduct the following test.
After removing data from the planetary anomaly, i.e., $8725 < {\rm HJD}^\prime < 8755$, we fit point-source point-lens (PSPL) models to the remaining light curve including the parallax effect.
This PSPL+parallax modeling is carried out by almost the same method with the standard modeling, which performs a grid search in the parallax parameter space $(\pien,\piee)$ and then additional modeling using local solutions.
From the grid search, we find only one local solution $(\pien,\piee)=(0.31,-0.67)$.
The modeling result shows that the point lens part of the light curve gives the usual 1D parallax constraint (see black and gray contours in Figure \ref{fig:pspl}).
We also add the result of the parallax+orbital model (colored contours) in Figure \ref{fig:pspl}.
This figure shows that the point lens parallax constraint is almost consistent with the constraint from fitting the full light curve including orbital motion within $3\sigma$, but shows that the planetary anomaly provides more precise information about parallax.

\begin{deluxetable}{lcccccccc}
\tablewidth{0pt}
\tablecaption{Degenerate models\label{tab-degen}}
\tablehead{
   Model                                      &       $1$              &&         $2$          &&      $3$          && $4$     
 }
\startdata
$\chi^2$                                     &  $4980.05$     &&   $4982.07$    &&  $4982.37$  && $4983.93$ \\
$\pien$                                       &  $0.45$            &&   $-0.07$         &&  $0.18$        && $0.52$ \\
$\piee$                                       &  $-0.40$          &&   $-0.20$         &&  $-0.30$      && $-0.35$ \\
$ds/dt$                                      &  $0.92$            &&   $-0.23$         &&  $-0.56$      && $-0.78$ \\
$d\alpha/dt$                              &  $0.01$            &&    $-0.54$         &&  $-0.11$       && $0.41$ \\
\hline
$M_{\rm h}$ $(M_\odot)$        &   $0.10$          &&   $0.15$           && $0.14$         && $0.07$ \\
$M_{\rm p}\, (M_{\rm J})$        &   $0.28$          &&   $0.99$          && $0.59$          && $0.33$ \\
$D_{\rm L}$ (kpc)                     &  $2.42$          &&   $3.83$          && $3.73$          && $2.69$ \\
\enddata
\tablecomments{The physical parameters were estimated with $(\thetae, \pie)$, where $\thetae$ is described in Section 4.}
\end{deluxetable}

\begin{figure}[t!]
\centering
\includegraphics[width=0.5\textwidth]{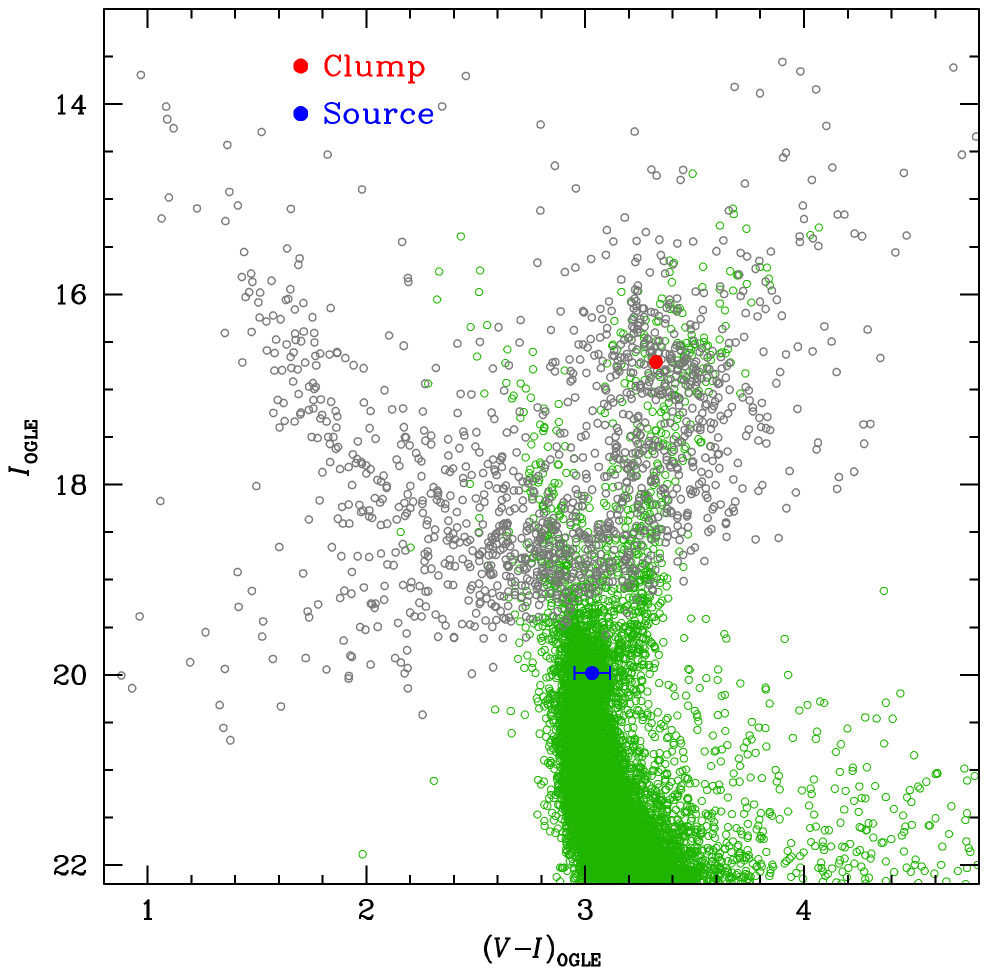}
\caption{CMD of stars around the event, which is constructed from combining OGLE and \textit{HST} observations.
The OGLE and \textit{HST} CMDs are plotted as gray and green open dots, respectively.
The red and blue solid dots indicate the positions of the red clump centroid and source, respectively.
\label{fig:cmd}}
\end{figure}

\section{Source Properties}

\begin{figure*}[t!]
\centering
\includegraphics[width=0.9\textwidth]{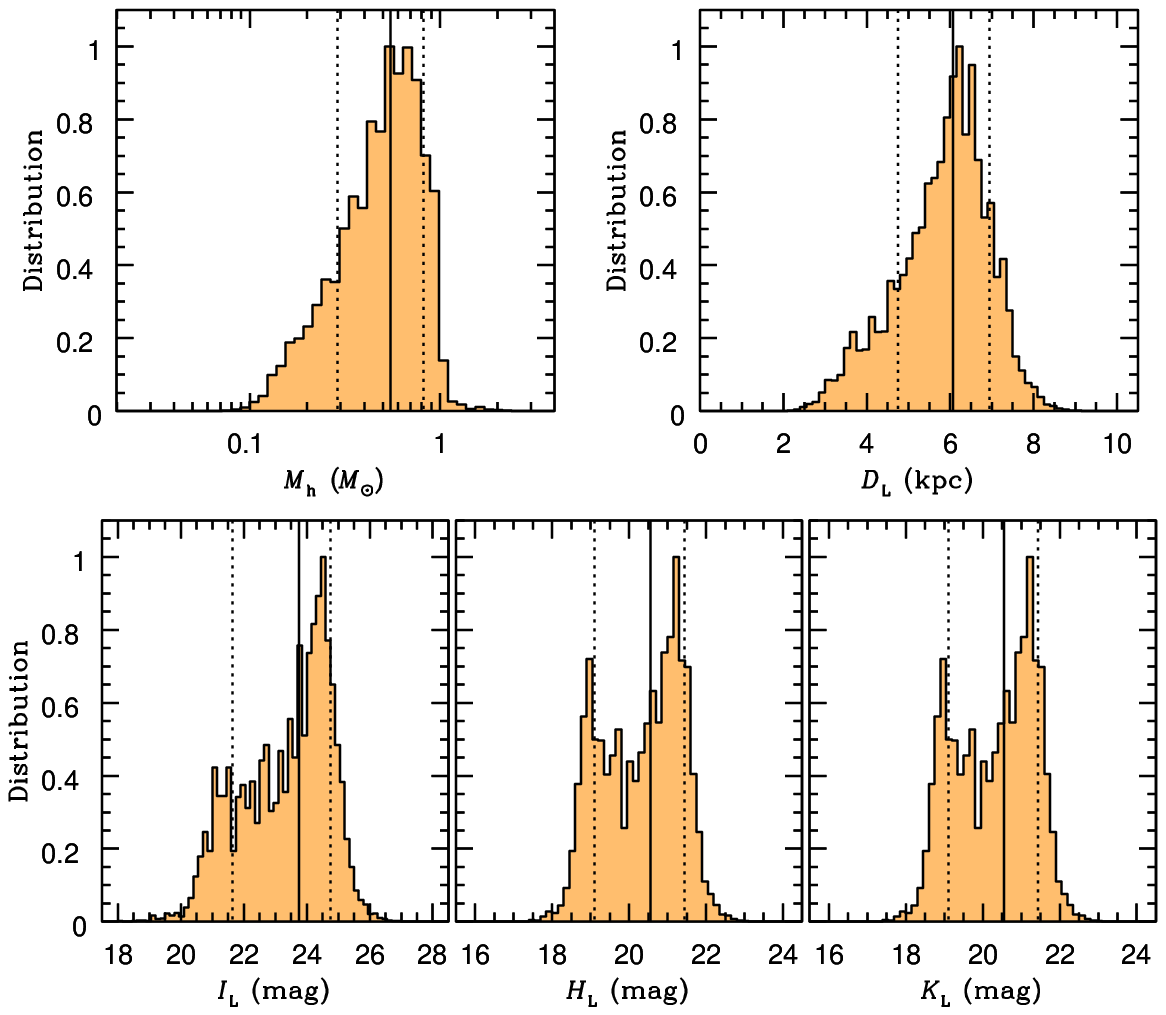}
\caption{Bayesian probability distributions of the mass, distance, and brightness of the host lens star.
The black solid vertical line and the two black dashed lines represent the median value and the 68\% confidence intervals of the distribution.
\label{fig:bayesian}}
\end{figure*}

As mentioned in Section 2, all KMT $V$-band data for the three sites were affected by bleeding.
We thus use the OGLE data sets for the source color and CMD.
From the OGLE CMD, we find that the color and magnitude of the clump are $(V-I, I)_{\rm cl}=(3.32, 16.71)$.
The brightness of the source is $I_{\rm s}=19.98 \pm 0.01$, which is obtained from the source flux of the best-fit model.
OGLE has three well-magnified $V$-band data points, but it turned out that they are not enough to get a precise source color.
We thus estimate the source color by combining the OGLE CMD and a CMD constructed from \textit{Hubble Space Telescope} (\textit{HST}) observations of Baade's window \citep{holtzman+1998}.
We note that the two CMDs are combined by calibrating the clump positions on each CMD, in which the the color and magnitude of the clump on the \textit{HST} CMD $(V-I, I)_{\it{HST},\rm cl}=(1.62, 15.15)$ is used from \citet{bennett+2008}.
We then extract \textit{HST} stars that have similar magnitudes to the source star, in which they are in the ranges of $17.63 \leqslant I_0 \leqslant 17.65$.
We estimate the mean color of the \textit{HST} stars and the standard deviation of the color and then take them as the source color and its uncertainty, respectively. 
From this, it is found that the source color is $(V-I)_{\it{HST},\rm s}=1.33 \pm 0.08$.
The angular source radius is estimated from the intrinsic color and magnitude of the source, which are determined from
\begin{equation}
(V-I, I)_{\rm s,0} = (V-I, I)_{\rm cl,0} + \Delta (V-I, I),
\end{equation}
where $\Delta (V-I)=(V-I)_{\it{HST},\rm s} - (V-I)_{\it{HST},\rm cl}$ and $\Delta I = I_{\rm s} - I_{\rm cl}$.
We adopt the intrinsic color and magnitude of the clump $(V-I)_{\rm cl,0}=1.06$ and $I_{\rm cl,0}=14.37$ from \citet{bensby+2011} and \citet{nataf+2013}, respectively.
Hence, we find that the intrinsic color and magnitude of the source are $(V-I, I)_{\rm s,0} = (0.77 \pm 0.08, 17.64 \pm 0.01 )$, indicating that the source is a late G dwarf.
The combined CMD is presented in Figure \ref{fig:cmd}.
By adopting the $VIK$ color-color relation of \citet{bessell&brett1988} and the color-surface brightness of \citet{kervella+2004}, we determine an angular source radius of $\thetas = 0.98 \pm 0.09\, \rm \mu as$.
From the determined $\thetas$ and $\rho$, we measure the angular Einstein radius
\begin{equation}
\thetae = {\thetas\over{\rho}}=\left\lbrace 
\begin{array}{ll}
0.458 \pm 0.044\, \textrm{mas} & \textrm{$(u_{0} > 0)$} \\
0. 443 \pm 0.042\, \textrm{mas} & \textrm{($u_{0} < 0)$}.
\end{array} \right.
\end{equation}

Then the relative lens-source proper motion is estimated as
\begin{equation}
\murel = {\thetae\over{\te}} = \left\lbrace 
\begin{array}{ll}
6.01 \pm 0.58\, \textrm{mas\, yr$^{-1}$} & \textrm{($u_{0} > 0)$} \\
5.86 \pm 0.55\, \textrm{mas\, yr$^{-1}$} & \textrm{$(u_{0} < 0)$}.
\end{array} \right.
\end{equation}

\section{Lens Properties}

As mentioned in Section 1, the physical lens parameters including the lens mass and distance to the lens are directly determined by two observables, $\thetae$ and $\pie$, which are defined as 
\begin{equation}
\label{eqn:mass}
M_{\rm L} =  {\thetae\over{\kappa \pie}};\quad \dl = {{\rm au}\over{\pie\thetae + \pi_{\rm S}}},
\end{equation}
where $\kappa \equiv 4G/(c^{2}\rm au) \approx 8.14\,{\rm mas}\, {M_\odot}^{-1}$ and $\pi_{\rm S}={\rm au}\, {\ds}^{-1}$ denotes the parallax of the source.
We adopt $\ds =7.73\, \rm kpc$ and $\pi_{\rm S}=0.13\, \rm mas$ in this work.
For \thisevent,~$\pie$ and $\thetae$ were measured, but $\pie$ was not constrained well.
We thus estimate the physical lens parameters by conducting a Bayesian analysis with the measured three observables of $(\te,\thetae,\pie)$ and the Galactic model of \citet{jung+2021}.
The Bayesian analysis assumes that all stars have an equal probability to host a planet with the observed mass ratio. 
For the Bayesian analysis, we first randomly generate $2 \times 10^7$ artificial microlensing events.
We then calculate the probability distributions of the physical lens parameters for events with $(\te,\thetae,\pie)$ located within the uncertainties of the three observables.

In order to estimate the lens brightness, we consider the extinction at a given lens distance.
According to \citet{bennett+2020}, the extinction to the lens, $A_{i,\rm L}$, is computed by 
\begin{equation}
A_{i,\rm L} = {1-e^{-|\dl({\rm sin}\ b)/h_{\rm dust}|}\over{1-e^{-|\ds({\rm sin}\ b)/h_{\rm dust}|}}}\ A_{i},
\end{equation}
where the index $i$ denotes the passband: $V$, $I$, $H$, or $K$; the dust scale height is $h_{\rm dust} = 120\ \rm pc$, and $A_{i}$ is the extinction to the source.
Using the information on the color and magnitude of the clump discussed in Section 4, we find $A_{I} = 2.34$ and $A_{V}=4.61$.
For the extinctions in the $H$ and $K$ bands, we adopt $A_{H} = 0.87$ and  $A_{K} = 0.53$, respectively, using the extinction law of \citet{cardelli+1989} for $R_{V}=3.1$,  i.e., $A_{H} = 0.190A_{V}$ and $A_{K} = 0.114A_{V}$.

Figure \ref{fig:bayesian} shows the probability distributions of the physical lens parameters estimated from the Bayesian analysis.
The physical lens parameters and their uncertainties represent the median values and 68\% confidence intervals of each distribution.
The mass and distance of the host star are estimated as
\begin{equation}
 M_{\rm h}=0.549^{+0.272}_{-0.260}\, M_\odot, \quad \dl=6.07^{+0.87}_{-1.32}\, \rm kpc.
\end{equation}
Then, the planet mass is determined as
\begin{equation}
 M_{\rm p}= qM_{\rm h} = 1.747^{+0.527}_{-0.507}\, M_{\rm J}.
\end{equation}
The projected star-planet separation is $a_\perp=5.19^{+0.90}_{-1.23}\, \rm au$.
According to $a_{\rm snow}=2.7 M/M_\odot$ \citep{kennedy&kenyon2008}, the snow line of the host is $a_{\rm snow} = 1.48^{+0.74}_{-0.70}\, \rm au$, indicating that the planet is orbiting beyond the snow line of the M dwarf star.
However, the host star could be also a K or a G dwarf star.

The probability distributions of the brightness of the lens star are shown in the bottom panel of Figure \ref{fig:bayesian}.
The brightness of the lens star is $I_{\rm L}=23.75^{+1.01}_{-2.11}$, $H_{\rm L}=20.55^{+0.89}_{-1.45}$, and $K_{\rm L}=20.05^{+0.84}_{-1.35}$.
The lens is 32 times fainter than the G dwarf source star of $I=19.98$, implying that it is possible to resolve them by follow-up observations.
Considering the relative lens-source proper motion $\murel = 6\, \rm mas\, yr^{-1}$, the lens will be separated from the source by $60\, \rm mas$ in 2029.
The separation $60\, \rm mas$ is about 4 times the FWHM at 1.6 $\mu\rm m$ for a next-generation 30 m telescope, e.g., the Giant Magellan Telescope (GMT; \citet{mcgregor+2012}) that will be operated in $\sim 2029$, thus the lens can be easily resolved with a 30 m telescope.

In addition,  \citet{vandorou+2023} have very recently reported the results of follow-up observations using Keck for OGLE-2016-BLG-1195.
The results show that a star with $K=20.0$, which is about 15 times fainter than the nearby star of $K=17.0$ at a separation of $56.4\, \rm mas$, can be resolved with Keck.
For this event, the brightnesses of the lens and source stars are $K_{\rm L}=20.1$ and $K=17.3$ and their separation is $60\,\rm mas$, thus their results suggest that this lens star can be detected with Keck.
If the lens flux could be detected, one can measure more precise lens properties, and then the orbital motion of the wide lens system can be better constrained.

\begin{deluxetable}{lccc}
\tablewidth{0pt}
\tablecaption{Physical lens parameters\label{tab-lens}}
\tablehead{
      Parameter                            &&&                   
 }
\startdata
$M_{\rm h}$ $(M_\odot)$        &&&  ${0.545^{+0.272}_{-0.260}}$  \\
$M_{\rm p}\, (M_{\rm J})$         &&&  ${1.747^{+0.527}_{-0.507}}$    \\
$D_{\rm L}$ (kpc)                     &&&  ${6.07^{+0.87}_{-1.32}}$         \\
$a_{\perp}$ (au)                       &&&  ${5.19^{+0.90}_{-1.23}}$          \\
$I_{\rm L}$ (mag)                      &&&  $23.75^{+1.01}_{-2.11}$     \\
$H_{\rm L}$ (mag)                    &&&  $20.55^{+0.89}_{-1.45}$   \\
$K_{\rm L}$ (mag)                    &&&  $20.05^{+0.84}_{-1.35}$    \\
$\thetae$ (mas)                        &&&  $0.46 \pm 0.04$                    \\  
$\murel\, (\rm mas\ yr^{-1})$    &&&  $6.01 \pm 0.58$  \\
\enddata
\end{deluxetable}

\section{Wide-orbit Planets}

\begin{figure*}[t!]
\centering
\includegraphics[width=0.8\textwidth]{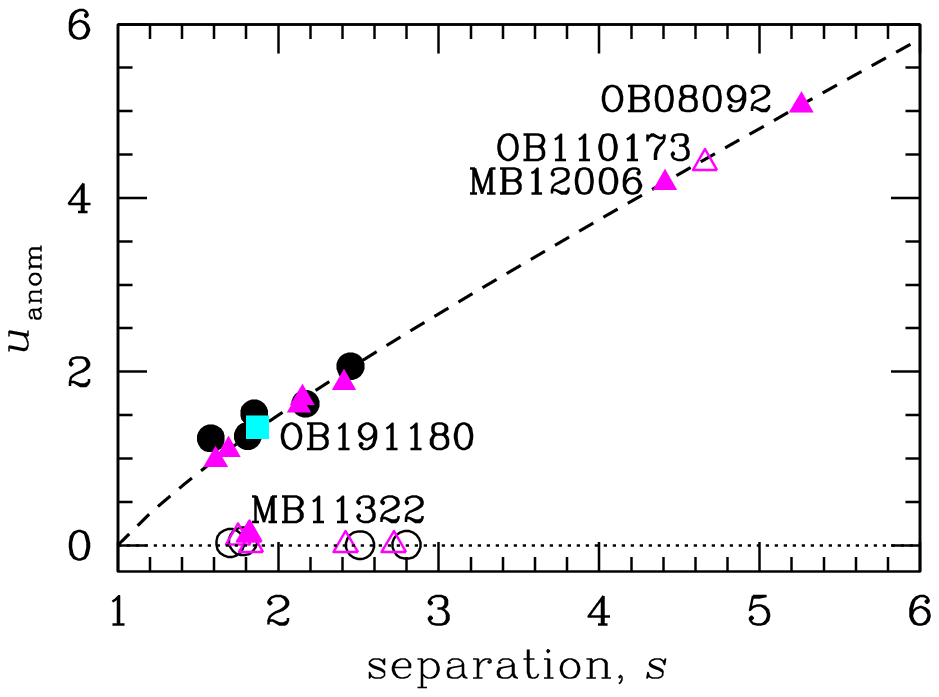}
\caption{Planets with $s > 1.5$ light-curve solutions from the 2018 and 2019 KMTNet seasons (black circles) and from data taken prior to 2015 (magenta triangles). 
Filled symbols are for planets with clear wide-orbit solutions; open symbols are for planets with $s < 1$ degeneracies. 
\thisevent Lb is shown as a square in cyan. Events noted in the text are labeled. 
The dotted black line is at $u_{\rm anom} = 0$. The dashed black line shows $u_{\rm anom} = s - 1/s$.
\label{fig:uanom}}
\end{figure*}

\begin{deluxetable*}{lcccccccl}
\tablecaption{Wide-orbit Planets\label{tab-wops}}
\tablehead{
\colhead{Name}                && \colhead{$\log q$} && \colhead{$s$} && \colhead{$u_{\rm anom}$} && \colhead{Reference(s)}
}
\startdata
Literature (pre-2015): \\
MOA-2007-BLG-400Lb    && -2.64    && 2.72     && 0.005    && \citet{dong+2009,bhattacharya+2021}\\
MOA-2011-BLG-028Lb     && -3.90    && 1.69    && 1.098     && \citet{skowron+2016} \\
MOA-2011-BLG-293Lb     && -2.28    && 1.83    && 0.004    && \citet{yee+2012} \\
MOA-2011-BLG-322Lb     && -1.55    && 1.82     && 0.126     && \citet{shvartzvald+2014} \\
MOA-2012-BLG-006Lb     && -1.78    && 4.41     && 4.170     && \citet{poleski+2017} \\
MOA-2013-BLG-605Lb     && -3.42   && 2.41     && 1.866    && \citet{sumi+2016} \\
MOA-bin-1Lb                       && -2.34  && 2.13      && 1.610     && \citet{bennett+2012} \\
MOA-bin-29b                      && -2.21   && 1.75     && 0.089    && \citet{kondo+2019} \\
OGLE-2005-BLG-390Lb    && -4.12   && 1.61     && 0.981     && \citet{beaulieu+2006} \\
OGLE-2008-BLG-092Lb    && -3.62  && 5.26    && 5.064    && \citet{poleski+2014} \\
OGLE-2011-BLG-0173Lb    && -3.34  && 4.66   && 4.403     && \citet{poleski+2018}\\
OGLE-2012-BLG-0563Lb  && -2.96   && 2.42    && 0.002    && \citet{fukui+2015} \\
OGLE-2012-BLG-0838Lb  && -3.40  && 2.15     && 1.696     && \citet{poleski+2020}\\
\hline
KMTNet (2018-2019): \\
KMT-2018-BLG-0030Lb    && -2.56  && 1.58   && 1.231      && \citet{jung+2022}\\ 
OGLE-2018-BLG-1367Lb   && -2.48  && 1.70    && 0.026     && \citet{gould+2022}\\ 
OGLE-2018-BLG-0383Lb  && -3.67  && 2.45   && 2.062      && \citet{wang+2022}\\ 
OGLE-2018-BLG-0567Lb  && -2.91   && 1.81    && 1.257      && \citet{jung+2021}\\ 
KMT-2019-BLG-0414Lb    && -2.26  && 2.80   && 0.005      &&\citet{han+2022}\\
KMT-2019-BLG-1953Lb    && -2.71   && 2.51    && 0.002     && \citet{han+2020a}  \\ 
OGLE-2019-BLG-1180Lb   && -2.57  && 1.87    && 1.359     && This work  \\ 
KMT-2019-BLG-0298Lb    && -2.53  && 1.85   && 1.520      && \citet{jung+2023}\\ 
OGLE-2019-BLG-0249Lb  && -2.11  && 1.78    && 0.046      && \citet{jung+2023} \\ 
OGLE-2019-BLG-0679Lb  && -2.41  && 2.17    && 1.630      && \citet{jung+2023} \\ 
\enddata
\end{deluxetable*}

\thisevent Lb is striking because it is a clear wide-separation planet detection with $s \sim 2$. 
By contrast, the core of microlensing's sensitivity to planets is $s \sim (0.62, 1.62)$. 
This may be calculated by assuming that in a typical planetary lensing event, the source position offset from the lens star is $u_{\rm anom} \le 1$ \citep{hwang+2022} and using the equation for the location of the planetary caustic 
\begin{equation}
u_{\rm anom} = s - 1/s
\label{eqn:u_anom}
\end{equation} 
from \citet{han2006}. 
By contrast, \thisevent~ has $u_{\rm anom}=1.36$, giving it a value of $s$ outside the standard ``lensing zone".
 
To place \thisevent Lb in context with known microlensing planets, we consider two samples of wide-orbit planets. 
First, we consider the microlensing planet discoveries from the systematic AF search of the 2018 and 2019 KMTNet seasons from which \thisevent\, is drawn \citep{gould+2022, jung+2022, jung+2023, zang+2022}. 
Second, we consider microlensing planets in the literature discovered in data prior to KMTNet (i.e., prior to 2015) taken from the NASA Exoplanet Archive (accessed 2023 May 1).
We limit this sample to planets with light-curve solutions that have $s > 1.5$ and $q < 0.03$ and only consider solutions with $\Delta \chi^2 < 10$ of the best fit. 
These planets are summarized in Table \ref{tab-wops}. 
Figure \ref{fig:uanom} shows $u_{\rm anom}$ vs. $s$ for these two samples of planets. 

The first feature of Figure \ref{fig:uanom} is that the planets are clearly delineated by $u_{\rm anom}$. 
The first group of planets are the planetary caustic anomalies, which have $u_{\rm anom} > 0.83$, which is what we would expect from Equation \ref{eqn:u_anom} given our limit $s > 1.5$. 
These planets all fall very close to the expected $u_{\rm anom}$ relation with some small scatter because source trajectory does not always pass through the exact center of the caustic. 
All but one of these planets has an unambiguous wide-orbit planet. 
The one exception is OGLE-2011-BLG-0173Lb \citep{poleski+2018}, which has an alternate, planetary caustic solution with $s < 1$ and a completely different value of $q$. 

In contrast to the planetary caustic events, the planets with $u_{\rm anom} \sim 0$ all (or likely all) suffer from close-wide degeneracy \citep{griest&safizadeh1998} due to being central caustic anomalies. 
MOA-2011-BLG-322Lb \citep{shvartzvald+2014} is the one possible exception. 
It has only an $s > 1$ solution in the literature, but the $s^{\dagger}$ analysis described in \citet{hwang+2022} and \citet{ryu+2022} reveals that in the corresponding $s < 1$ solution, the angle of the source trajectory is such that it passes near or through the planetary caustic, creating an extra signal that would nominally exclude such models. However, \citet{shvartzvald+2014} only considered static models; it seems likely that adding the parallax or orbital motion of the planet would allow for a plausible $s < 1$ solution that avoids the planetary caustic. 

For planets with $1.5 < s < 3$, KMTNet has a significant advantage over early microlensing detections: six out of 10 detections are planetary caustic detections. 
By contrast, the early microlensing detections had a much higher proportion of central caustic events (50\%). Most likely, this is due to the need for follow-up observations \citep{gould&loeb1992} and the subsequent bias toward high-magnification events \citep{udalski+2005}. 
Hence, KMTNet is fulfilling its promise to detect a larger number of planetary caustic events. 
This is essential for studying the dependence of planet occurrence on separation \citep{poleski+2021} because, as shown above, these are the events with a clear measurement of the host-planet separation.

However, the early microlensing detections also show a class of planets with $s > 4$ that have no KMTNet counterparts to date. 
They are OGLE-2008-BLG-092 \citep{poleski+2014}, OGLE-2011-BLG-0173 \citep{poleski+2018}, and MOA-2012-BLG-006 \citep{poleski+2017}. 
In these cases, the planetary anomalies all occurred near the beginning or end of the observing season, with a separation from the peak that was, respectively, 67\%, 52\%, and 28\% of the total duration of the observing season. 
Hence, it is possible that similar anomalies occur for KMTNet events, but they might fall during gaps between the observing seasons or even in other seasons entirely, e.g., if the peak of the stellar event is shifted with respect to the midpoint of the season. 

To explore further the possibility of missed planets with KMTNet, we consider the expected ratios of central caustic to planetary caustic events. 
The probability that a planet is detected through a caustic crossing is proportional to the size of that caustic. 
\citet{chung+2005} and \citet{han2006} give approximations for the sizes of the central and planetary caustics, respectively
\begin{equation}
\Delta \xi_{\rm cent} \sim \frac{4 q}{(s - 1/s)^2} ; \quad 
\Delta \xi_{\rm pl} \sim \left(\frac{4\sqrt{q}}{s^2}\right)\left(1 + \frac{1}{2s^2}\right) .
\label{eqn:xi_ratio}
\end{equation}
Hence, the ratio $\Delta \xi_{\rm pl} / \Delta \xi_{\rm cent}$ scales as $q^{-1/2}$ and, in the limit $s \gg 1$, reduces to $q^{-1/2}$. 
So, for $\log q = -2.5$, we expect $\sim 18$ planetary caustic-type events for every central caustic event as $s \rightarrow \inf$. 

For the KMTNet sample, $\log q = -2.5$ and $s = 2$ are typical values. 
This suggests that there should be $\sim 11$ planetary caustic events (either with or without host star detections) for every central caustic event. 
This is much larger than our observed ratio 6:4. 
Of course, this does not take into account several factors. 
First, not all planetary caustic events will have a detectable host star. 
Some of the 11 might manifest as free-floating planet candidates, and so might be excluded from the AF search. 
Alternatively, some of the hosts may not manifest as a separate peak, but only as a distortion to the planetary event, such as in MOA-bin-1 \citep{bennett+2012}, and there may be separate detection effects for distorted, short-timescale events. 
This simple calculation also does not take into account any observability criteria such as observing window (as discussed above) or signal-to-noise (which creates a bias toward central caustic perturbations because of their higher magnifications).
Finally, some or all of the central caustic events could be due to planets with a separation of $s^{-1}$; disentangling this contribution would require knowledge of the underlying separation distribution of planets.
However, given that there are $s > 4$ planets from the literature that so far have no counterparts in KMTNet data, it would be worthwhile to consider more carefully whether additional wide-orbit planets might be missing from the KMTNet sample. In particular, a search for planetary anomalies in data from observing seasons adjacent to the main stellar event could yield additional planets.

\section{Conclusion}
We analyzed the planetary lensing event OGLE-2019-BLG-1180, which has remarkable anomalies near the baseline after the peak of the light curve.
We estimated the physical lens parameters by conducting a Bayesian analysis using the measured observables of $(\te, \thetae, \pie)$.
From the Bayesian analysis, it was found that the lens system is composed of a late-type star of $0.55^{+0.27}_{-0.26}\, \rm M_\odot$ and a super-Jupiter-mass planet of $1.75^{+0.53}_{-0.51}\, \rm M_{\rm J}$ at a distance $\dl =6.1^{+0.9}_{-1.3}\, \rm kpc$.
The projected star-planet separation is $a_\perp=5.2^{+0.9}_{-1.2}\,\rm au$, which indicates that the planet lies beyond the snow line of the host star.
Considering $\murel = 6\, \rm mas\, yr^{-1}$, the lens flux can be resolved by adaptive optics of Keck or a next-generation 30 m class telescope in the future.

\section{Acknowlegement}
The work by S.-J.C. was supported by the Korea Astronomy and Space Science Institute under the R\&D program (project No. 2023-1-832-03) supervised by the Ministry of Science and ICT.
J.C.Y., I.-G.S., and S.-J.C. acknowledge support from N.S.F grant No. AST-2108414.
The work by C.H. was supported by the grants of National Research Foundation of Korea (2020R1A4A2002885 and 2019R1A2C2085965).
This research has made use of the KMTNet system operated by the KASI and the data were obtained at three sites of CTIO in Chile, SAAO in South Africa, and SSO in Australia. 
Data transfer from the host site to KASI was supported by the Korea Research Environment Open NETwork (KREONET).

\end{document}